\documentclass[aps,prd,reprint,superscriptaddress]{revtex4-2}
\usepackage[english]{babel}
\usepackage{lmodern}
\usepackage{amsmath,amsfonts,amssymb}
\usepackage[autostyle]{csquotes}
\usepackage{graphicx}
\usepackage{dcolumn}
\usepackage{bm}
\usepackage[table,xcdraw]{xcolor}
\usepackage{caption}
\usepackage{placeins}
\usepackage[format=plain]{caption}
\usepackage{hyperref}
\usepackage{float}
\usepackage[belowskip=-15pt,aboveskip=11pt]{caption}
\newcommand{\Driving}{{\mathchoice{}{}{\scriptscriptstyle}{}D}}
\newcommand{\Superconducting}{{\mathchoice{}{}{\scriptscriptstyle}{} S}}
\newcommand{\Normalconducting}{{\mathchoice{}{}{\scriptscriptstyle}{} N}}

\hypersetup{
  colorlinks=true,
  citecolor=blue,
  linkcolor=blue,
  urlcolor=blue}
  \makeatletter

\begin{document}
\makeatletter
    \renewcommand\@make@capt@title[2]{%
     \@ifx@empty\float@link{\@firstofone}{\expandafter\href\expandafter{\float@link}}%
      {\textbf{#1}}\@caption@fignum@sep#2\quad}%
    \makeatother
\title{Mixed\,-State Magnetotransport Below T$_c$ in Type\,-II Superconducting NbN Thin Films } 

\author{Sai Vaishnavi Kanduri}
\altaffiliation{\href{mailto:sai.lanka@uni-jena.de}{sai.lanka@uni-jena.de}}
\affiliation{Leibniz Institute of Photonic Technology,\,07745 Jena,\,Germany}
\affiliation{Institute for Solid State Physics,\,Friedrich Schiller University Jena,\,07743 Jena,\,Germany} 
\author{Danilo Bürger}
\affiliation{Helmholtz-Zentrum Dresden-Rossendorf (HZDR) Innovation GmbH,\,Bautzner Landstraße 400,\,01328 Dresden,\,Germany}
\author{Zichao Li}
\affiliation{Institute of Ion-Beam Physics and Materials Research, Helmholtz-Zentrum Dresden-Rossendorf\,(HZDR),\,Bautzner Landstraße 400,\,01328 Dresden,\,Germany}
\author{Mario Ziegler}
\affiliation{Leibniz Institute of Photonic Technology,\,07745 Jena,\,Germany}
\author{Detlef Born}
\affiliation{Leibniz Institute of Photonic Technology,\,07745 Jena,\,Germany}
\author{Sahitya\,V.\,Vegesna}
\altaffiliation{\href{mailto:sahitya.vegesna@uni-jena.de}{sahitya.vegesna@uni-jena.de}}
\affiliation{Leibniz Institute of Photonic Technology,\,07745 Jena,\,Germany}
\affiliation{Institute for Solid State Physics,\,Friedrich Schiller University Jena,\,07743 Jena,\,Germany}
\author{Heidemarie Krüger}
\altaffiliation{\href{mailto:heidemarie.krueger@uni-jena.de}{heidemarie.krueger@uni-jena.de}}
\affiliation{Leibniz Institute of Photonic Technology,\,07745 Jena,\,Germany}
\affiliation{Institute for Solid State Physics,\,Friedrich Schiller University Jena,\,07743 Jena,\,Germany}
\date{\today}
\begin{abstract}
The transport properties of type\,-II superconductors in the mixed state below the superconducting transition temperature, T$_{c}$, remain difficult to interpret because superconducting and dissipative processes coexist in the presence of a magnetic field. To investigate this regime in NbN thin films, we measured the field\,-dependent sheet resistance ($R^M_{xx}$) and Hall resistance ($R^M_{xy}$) below T$_{c}$ in Van der Pauw geometry for 4\,nm thick \textit{NbN} films on two different substrates, sapphire and silicon, at two driving currents, 1\,mA and 2\,mA. By combining the longitudinal response with the antisymmetrized Hall signal, we use a phenomenological current\,-partition model to estimate the dissipative and superconducting current contributions. From this analysis, we estimate the field-dependent superconducting current together with Hall-derived effective dissipative carrier density and mobility. We also observe an even-in-field transverse signal whose magnitude depends strongly on the driving current. We discuss this signal as a possible mixed-state response involving the magnetic and supercurrent texture of Abrikosov vortices, vortex dynamics, and Cooper-pair current redistribution. The present measurements do not uniquely distinguish vortex, condensate, or combined contributions. The model provides a practical way to compare the longitudinal and transverse mixed-state responses without implying a direct spatial separation of the two current channels.
\end{abstract}
\maketitle
\section{\label{sec:level1}Introduction}
Detecting single photons is a key requirement in numerous areas of science and technology, including quantum and fundamental research \cite{article1,article2}, optical communication \cite{article3,article4,article5}, long-range optical  sensing\cite{article6,article7,article8,article9}, spectroscopy \cite{article10,article11,article12} and astronomy \cite{article13}. The need to detect extremely weak photons has driven the development of several single photon-detectors, such as silicon photomultipliers (SiPMs) \cite{article14}, single photon avalanche detectors (SPADs) \cite{article15}, and superconducting nanowire single-photon detectors (SNSPDs) \cite{article16}, each offering advantages in specific wavelength ranges. Among these SNSPDs  emerged as a leading technology  because of high detection efficiency, low dark-count rate, low timing jitter, fast reset time and broadband sensitivity \cite{article17,article18}. Recent advancements have pushed SNSPD performance to near unity detection efficiency in the near infrared and telecom spectral range. Fiber-coupled SNSPDs have recorded system detection efficiencies of 98.0 $\pm$ 0.5 \% at 1550 nm and $99.5^{+0.5}_{-2.07}$ \% at 1350 nm, while waveguide-integrated SNSPDs have achieved efficiencies above 99\% at 1550 nm \cite{article19,article20,article21,article22}.

Despite the excellent performance of  SNSPDs in the visible, near infrared, and telecom regimes, the microscopic origin of the detection event remains incompletely understood. At the device level, SNSPD detection is described as a sequence of photon absorption, followed by local suppression of the superconducting state and formation of a dissipative region in the nanowire, and electrical detection of the resulting voltage pulse \cite{article17}. However, the physical origin of this dissipative region remains under active discussion, and is interpreted using hotspot-based models, while vortex-based models emphasize the possible role of vortex entry, motion and vortex assisted switching of nanowire into a resistive state \cite{article23,article24,article25,article26,article27}. 

To understand how a current biased superconducting nanowire evolves into a dissipative state, it is important to examine the superconducting properties of the underlying material. Type-II superconducting Niobium Nitride (\textit{NbN}) is one of the most widely used materials for SNSPDs because of its  high critical temperatures ($T_{c}$) of around 16\,K, high critical current densities ($J_{c}$) \cite{article17,article28}. These are of interest for a variety of superconducting devices, ranging from hot-electron bolometer mixers \cite{article29,article30} to micro resonators \cite{article31}, magnetometers \cite{article32} and SNSPDs \cite{article17,article33}. In these devices, the superconducting state is controlled by three main critical parameters: temperature, current density, and magnetic field. While SNSPD operation mainly relies on the current criticality, with the nanowire biased close to $J_{c}$, magnetic-field-dependent transport provides a controlled way to examine how vortex formation, field-induced dissipation, and the magnetic suppression of superconductivity in \textit{NbN} below $T_{c}$ occurs. Previous studies have used magnetic field to study vortex dynamics and tune the  performance  of \textit{NbN}-SNSPD through I-V characteristics and count rate measurements \cite{article25,article27, article34}. However, direct magnetotransport and Hall measurements below-$T_{c}$ in structured NbN films remain limited, particularly for understanding field-induced dissipation and vortex-related transport.

To interpret such field-induced transport, it is useful to consider how type-II \textit{NbN} responds when an external magnetic field $H_{o/p}$ is applied below $T_{c}$. The response evolves through three regimes defined by the lower and upper critical fields, $H_{c_{1}}$ and $H_{c_{2}}$. At low fields, the superconducting state dominates. At intermediate fields ($H_{c_1}<H_{o/p}<H_{c_{2}}$), magnetic flux penetrates the film as quantized Abrikosov vortices, producing a mixed state in which superconducting and dissipative transport can coexist \cite{Abrikosov1957}. Under an applied current, mobile vortices produce flux-flow dissipation and can contribute to a transverse voltage. A normal-core approximation is useful for describing the dissipative channel \cite{article47}, although the vortex Hall response can contain both core and hydrodynamic contributions \cite{article48}. At sufficiently high fields, superconductivity is suppressed and the transport becomes normal-state like \cite{article28}.

In our previous work, we investigated magneto transport in ALD-grown \textit{NbN} thin films in the normal-state regime above $T_{c}$ where magnetoresistance and Hall measurements were used to characterize electronic disorder, weak-localization behavior, interaction effects, carrier concentration, and mobility \cite{article35}. We further examined the transition regime around $T_{c}$, where longitudinal and Hall measurements revealed fluxoid-driven dissipation in the mixed-state and enabled the extraction of fluxoid-related quantities such as areal density, mobility, and dissipative current contributions \cite{article28}. The present work extends this framework into the superconducting state below-$T_{c}$. In particular, we examine the longitudinal and Hall magneto transport response of \textit{NbN} films under different drive currents, on different substrates, and in Van der Pauw and Hall-bar geometries.

\section{Experimental Details}
NbN thin films with different nominal thicknesses were deposited on substrates using plasma-enhanced atomic layer deposition (PEALD). The deposition process followed the previously reported PEALD growth procedure for NbN films, using (tert-butylimido)-tris(diethylamido)-niobium (TBTDEN) and hydrogen plasma as precursors, with the substrate temperature maintained at 380\,$^{\circ}$C \cite{article36,article37,article38}. The nominally 4\,nm samples were deposited using 72 cycles with a 3.3\,nm target thickness, whereas the 25\,nm samples were deposited using 543 cycles. The pulse sequence was 3000\,ms precursor dose, 20\,s purge, 60\,s H$_2$ plasma at 300\,W and 5\,sccm, followed by a 10\,s purge, with a reported growth per cycle of 0.46\,\text{\AA}/cycle.

The Van der Pauw measurements were performed on nominally 4\,nm thick NbN films deposited on Si and sapphire substrates. The 25\,nm NbN films on Si and glass were patterned into Hall bars by photolithography using an approximately 1.5\,$\mu$m AZ1514H resist, followed by CF$_4$/O$_2$ plasma etching in a capacitively coupled plasma system. The sample and measurement matrix is summarized in Table~\ref{tab:sample_overview}; the Hall-bar data and analysis are given in Supplementary Figs.~S1--S6.

\begin{table*}[t]
\caption{Samples and measurement conditions used in this work. The transition temperatures and dimensions are taken from the sample records and transport analysis.}
\label{tab:sample_overview}
\begin{ruledtabular}
\begin{tabular}{lccccccc}
Sample & Geometry & Substrate & $t$ (nm) & $T_c$ (K) & $T$ analysed (K) & $I_D$ (mA) & Field/data \\
\hline
4\,nm NbN/Si & Van der Pauw & Si & 4 & 8.60 & 5.0, 7.5 & 1, 2 & o/p $R_{xx},R_{xy}$; i/p $R_{xx}$ \\
4\,nm NbN/sapphire & Van der Pauw & sapphire & 4 & 8.77 & 5.0, 7.5 & 1, 2 & o/p $R_{xx},R_{xy}$; i/p $R_{xx}$ \\
25\,nm NbN/Si & Hall bar & Si & 25 & 12.54 & 8, 10 & 2 & o/p $R_{xx},R_{xy}$ \\
25\,nm NbN/glass & Hall bar & glass & 25 & 12.34 & 8, 10 & 2 & o/p $R_{xx},R_{xy}$ \\
\end{tabular}
\end{ruledtabular}
\end{table*}

The samples were mounted in a Lake Shore HMS 9700A Hall measurement system. The measurement temperatures were selected below the superconducting transition temperature $T_c$. A driving current $I_D$ was applied, and the resulting voltage response was recorded as a function of temperature and magnetic field. Measurements were performed in magnetic fields up to 6\,T. The longitudinal response was measured as sheet resistance ($R^M_{xx}$) for fields perpendicular and parallel to the film plane, whereas the transverse response ($R^M_{xy}$) was measured with the field perpendicular to the film. The Van der Pauw measurements were performed at 5.0\,K and 7.5\,K using $I_D=1$ and 2\,mA. The Hall-bar measurements at 8\,K and 10\,K used $I_D=2$\,mA and are reported in the Supplementary Material.

\setcounter{figure}{1}
\begin{figure}[H]
\begin{center}
\includegraphics[width=1.0\columnwidth]{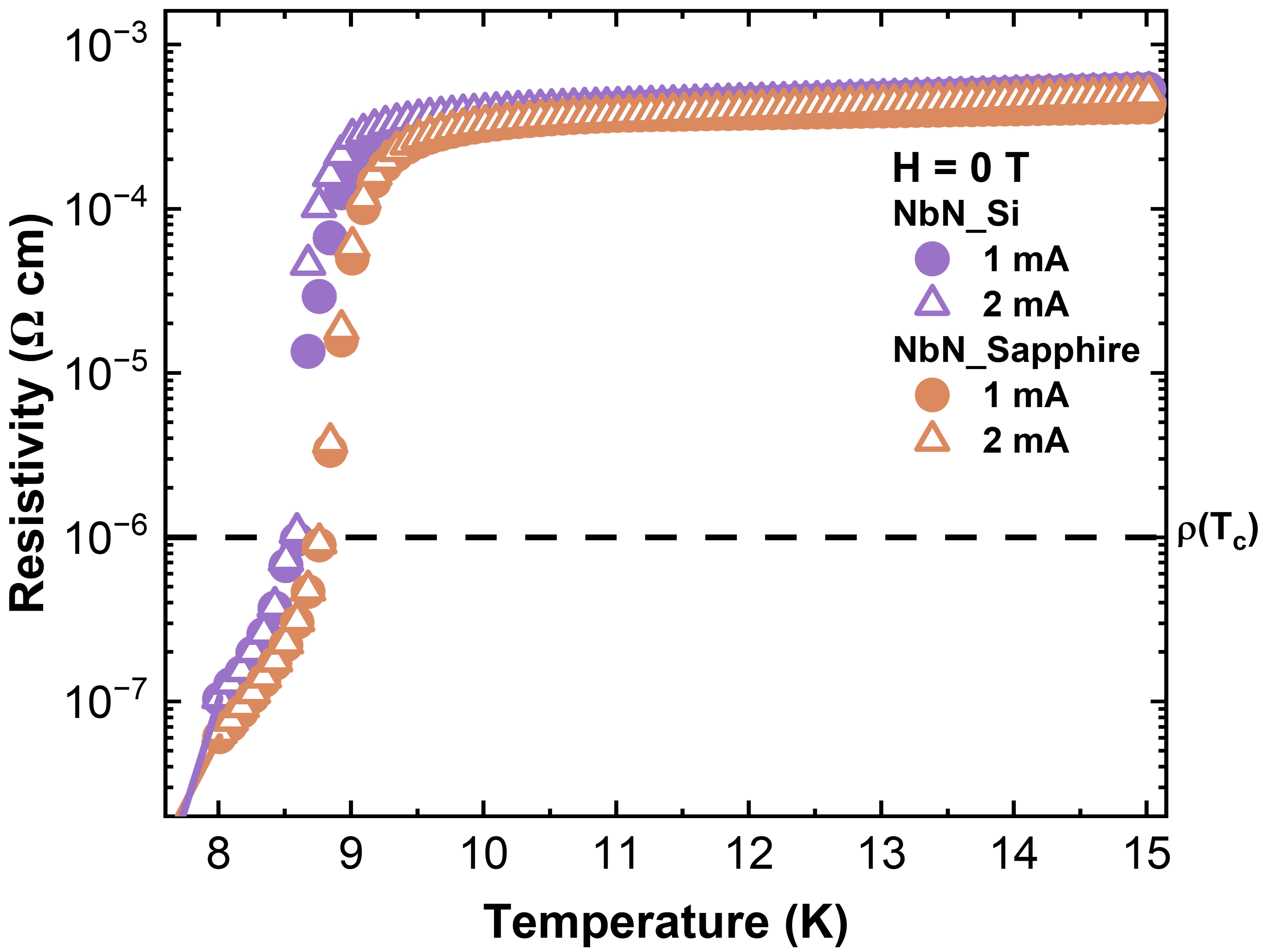}
\caption{Zero\,-field temperature-dependent resistivity, $\rho^M_{xx}(T)$, of 4\,nm thick \textit{NbN} thin films grown on Si and sapphire substrates in Van der Pauw geometry, measured at applied currents of 1\,mA and 2\,mA. The critical temperature was determined using the criterion $\rho^M_{xx}(T_c)\approx1\,\mu\Omega$cm. Solid lines shown below 8\,K represent extrapolated trends included for visual guidance only and are not experimental data.}
\label{fig:2}
\end{center}
\end{figure} 
\section{Results and Discussion}
The magneto\,-transport properties of the 4\,nm thick NbN thin films grown on Si and sapphire substrates were investigated using temperature- and field\,-dependent transport measurements in Van der Pauw geometry. In this section, the zero\,-field superconducting transition is discussed first to determine the critical temperature, followed by analysis of the field\,-dependent longitudinal and transverse transport behavior below T$_c$. Fig.~\ref{fig:2} shows the zero-magnetic-field temperature-dependent resistivity of 4\,nm thick \textit{NbN} thin films grown on silicon and sapphire substrates, measured in Van der Pauw geometry for applied currents of 1\,mA and 2\,mA. In all samples, the resistivity decreases strongly with decreasing temperature, indicating the onset of the superconducting transition in the absence of an external magnetic field. The overall transition behavior is similar for the two substrates, and for each substrate the curves obtained at 1\,mA and 2\,mA nearly overlap, indicating small current dependence within the investigated range. To consistently compare transition temperatures, the critical temperature was defined as the temperature at which the resistivity dropped to 1\,$\mu$$\Omega$cm. Using this criterion, we determined $T_{c}$ to be 8.60\,K for NbN on Si and 8.77\,K for NbN on sapphire. The slightly higher $T_{c}$ obtained for NbN on sapphire suggests a substrate\,-dependent influence on the superconducting properties of the ultrathin film. As the lowest measured temperature was 8\,K, the resistive transition below this point could not be resolved. Therefore, the solid lines shown below 8\,K are included only as visual guides and do not represent the measured data. Similar behavior has previously been reported for ultrathin NbN films, where a limited temperature range and residual stray magnetic fields were suggested as possible contributing factors \cite{article35}.
\begin{figure}[t!]
\begin{center}
\includegraphics[width=1.0\columnwidth]{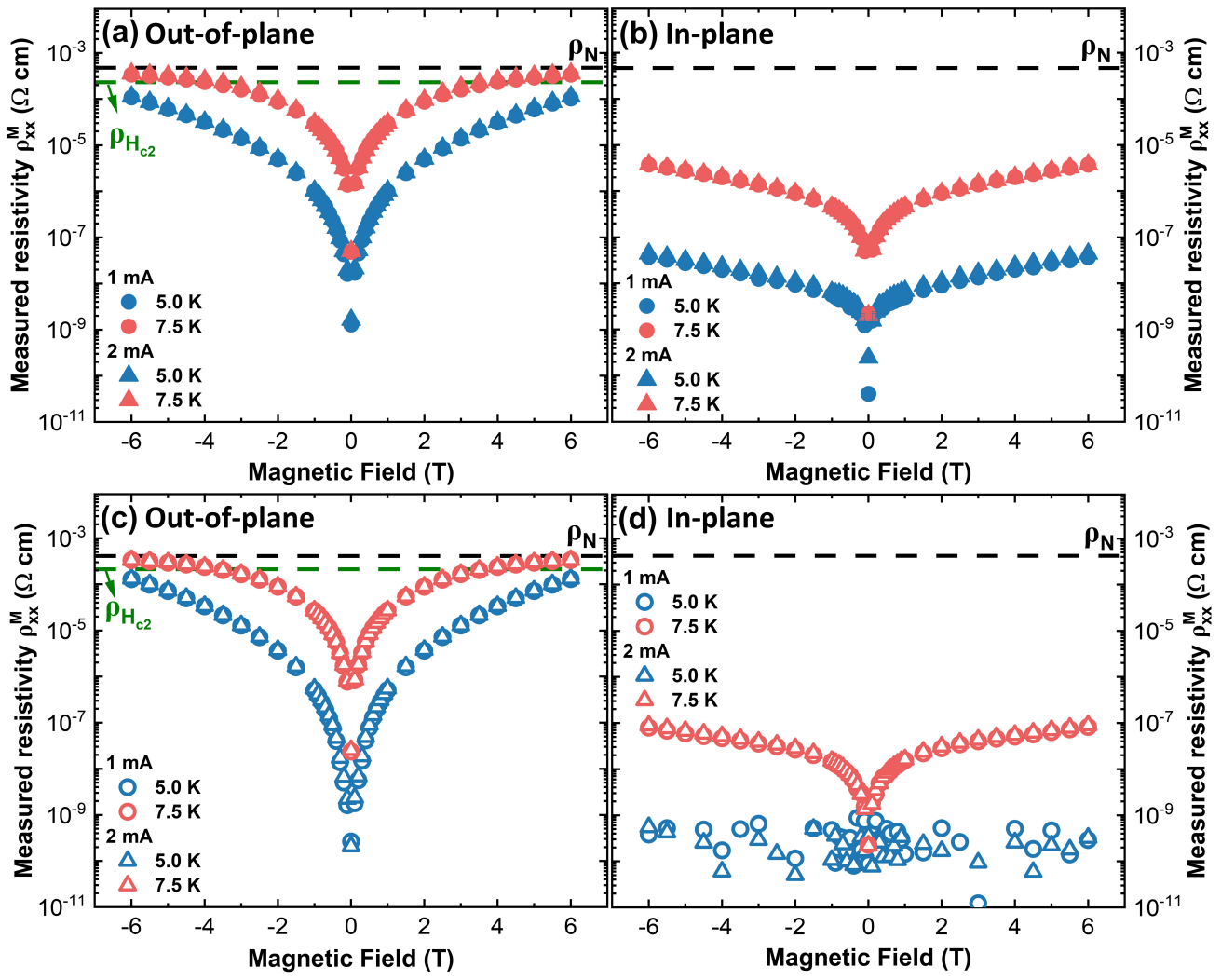}
\caption{Longitudinal resistivity, $\rho^M_{xx}$, as a function of magnetic field in out-of-plane and in-plane configurations at 5.0\,K and 7.5\,K below T$_{c}$ for 4\,nm thick \textit{NbN} thin films on (a,b) Si (closed symbols) and (c,d) sapphire (open symbols). Measurements were performed from $-6$\,T to $+6$\,T with driving currents of 1\,mA and 2\,mA. At higher fields, the measured resistivity (marked in black) approaches the normal-state \textit{NbN} resistivity, taken as approximately 463\,$\mu$$\Omega$cm for NbN/Si and 423\,$\mu$$\Omega$cm for NbN/sapphire.}
\label{fig:3}
\end{center}
\end{figure} 
\subsection{Longitudinal magnetotransport $\rho^M_{xx}$}
To investigate the effect of magnetic field on the superconducting state below T$_{c}$, field-dependent longitudinal resistance measurements were performed on NbN thin films in Van der Pauw geometry, using both out-of-plane and in-plane configurations. Fig.~\ref{fig:3} shows the magnetic-field dependence of the longitudinal resistivity, $\rho^M_{xx}=R^M_{xx}t$. For both substrates and field orientations, the resistivity increases with field, indicating the gradual suppression of superconductivity. The upper characteristic field, denoted $H_{c_2}$ in this transport analysis, was identified using the 50\% criterion at which $\rho^M_{xx}$ reaches one-half of the normal-state reference resistivity, $\rho_N/2$. From the tabulated analysis associated with Appendix Fig.~\ref{fig:7}, the 7.5\,K values are approximately $|H_{c_2}|=3.49$\,T for NbN/Si and $|H_{c_2}|\approx2$--3\,T for NbN/sapphire for both currents. At 5\,K, the criterion is not reached within the $\pm6$\,T measurement range, so only $|H_{c_2}|\geq6$\,T is reported. These are transport-defined crossover fields rather than independent thermodynamic determinations of $H_{c_2}$. The Hall-bar longitudinal response and its characteristic-field analysis are shown in Supplementary Figs.~S1 and S2.

To further identify the field range over which the transport response changes most strongly, we analysed the field derivative of $R^M_{xx}$ in Appendix Fig.~\ref{fig:7}. The slope change is used as a transport onset of measurable field-induced dissipation. The analysis gives $|H_{c_1}|\approx1.49$\,T at 5\,K and $|H_{c_1}|\approx0.09$\,T at 7.5\,K for both substrates and currents, as summarized in Table~\ref{tab:Hc_values}. Together with the 50\% crossover fields denoted $H_{c_2}$, these values define the mixed-state transport window used in the discussion. Because demagnetization, finite thickness, pinning, and the resistive criterion can shift the onset, $H_{c_1}$ is used here as a transport-defined characteristic field and not as a direct thermodynamic lower critical field.

A clear anisotropy is observed between the out-of-plane and in-plane magnetic field orientations. When the magnetic field is applied perpendicular to the film, the rise in resistivity is much stronger than in the parallel field case. This indicates that the superconducting state is suppressed more efficiently by a perpendicular magnetic field. In this geometry, the field can drive the film into a mixed-state regime, where vortex/fluxoid motion under the applied current can produce finite dissipation. With increasing field, the superconducting contribution is progressively reduced and the transport response evolves toward the normal-conducting state. In contrast, the in\,-plane configuration shows a weaker, though still finite, resistive response. This indicates that the superconducting state is less strongly disturbed when the magnetic field is applied parallel to the film. The residual in\,-plane dissipation may arise from slight field misalignment and edge-related flux-entry or vortex processes \cite{book46,article47}. Overall, the longitudinal magnetotransport identifies the field range over which superconducting and dissipative transport coexist, providing the basis for the subsequent analysis of the superconducting current.

\subsection{Analysis of the superconducting current}
To further quantify the coexistence of superconducting and dissipative transport below T$_{c}$, we analyse the measured longitudinal resistivity in terms of an estimated superconducting-current contribution, $I_{\Superconducting}$, using the same phenomenological current\,-partition framework adopted in our earlier work around T$_{c}$ \cite{article28}. In this approach, the total driving current is written as follows: 
\begin{eqnarray}
I_{\Driving} = I_{\Superconducting}(H)+I_{\Normalconducting}(H) ,
\label{Eq.1}
\end{eqnarray}
where $I_{\Superconducting}$ and $I_{\Normalconducting}$ denote the superconducting and effective dissipative contributions within the model. The subscripts D, S, N, and M symbolize driving, superconducting, normal conducting, and measured, respectively. The measured longitudinal voltage is attributed to the dissipative contribution, which allows the superconducting contribution to be estimated from the field-dependent resistivity by comparison with a suitable normal-state reference. This is a transport-based partition and should not be interpreted as a direct spatial separation of two independently measured currents. In the previous analysis around T$_{c}$, the high-field resistivity at each temperature was used as the normal-conducting resistivity, because the superconducting contribution is weak in the transition region and the applied magnetic field can drive the system close to the normal state. However, in the present below-T$_{c}$ measurements, the superconducting state remains finite over the available magnetic-field range. Therefore, the high-field value of the measured resistivity cannot be used as the normal-state resistivity, because it may still contain a superconducting contribution.

To avoid underestimating the normal-state resistivity, we use the zero-field resistivity measured at 15 K as a normal-state reference value, since this temperature lies above the critical temperature, T$_{c}$. 

\begin{eqnarray}
 \rho^{ref}_{N} = \rho^M_{xx}(0,15\,\mathrm{K})
   \label{Eq.2}
\end{eqnarray}

Using this reference the superconducting current is estimated as follows:

\begin{eqnarray}
   I_{\Superconducting}(H,T) = I_{\Driving}\left(1-\frac{\rho^M_{xx}(H,T)}{\rho^{ref}_{N}}\right).
   \label{Eq.3}
\end{eqnarray}

\begin{figure}[t!]
\begin{center}
\includegraphics[width=1.0\columnwidth]{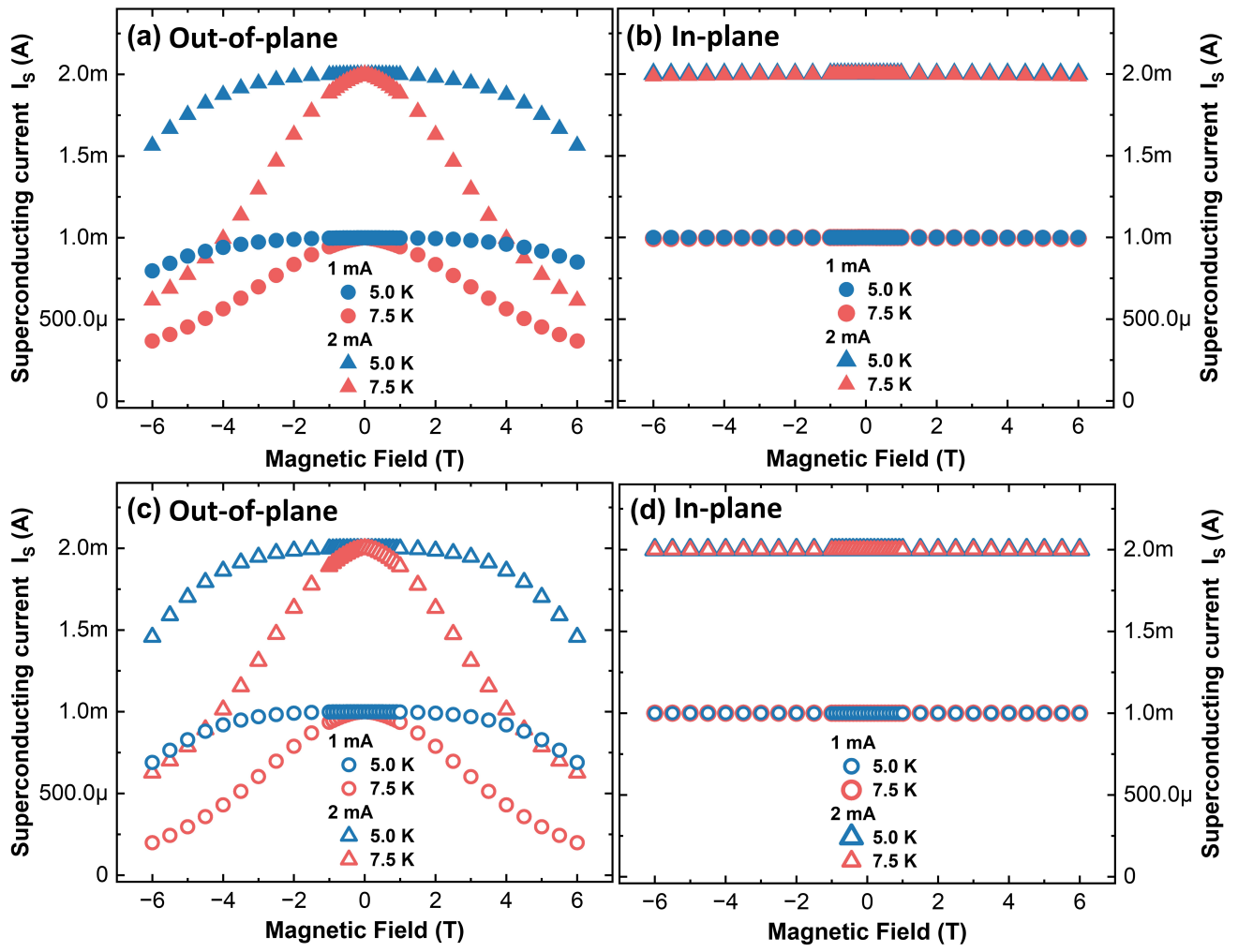}
\caption{Estimated superconducting-current contribution, $I_{\Superconducting}$, obtained from the measured longitudinal resistivity, $\rho^M_{xx}$, using the phenomenological partition in Eq.~(\ref{Eq.3}) for 4\,nm thick \textit{NbN} films on (a,b) Si (closed symbols) and (c,d) sapphire (open symbols), plotted as a function of out-of-plane and in-plane magnetic field at 5.0\,K and 7.5\,K. Measurements were carried out from $-6$\,T to $+6$\,T with applied currents of 1\,mA and 2\,mA. For an out-of-plane magnetic field, $I_{\Superconducting}$ decreases with increasing temperature and magnetic field. In contrast, under an in-plane magnetic field, $I_{\Superconducting}$ remains close to the applied driving current throughout the measured field range, indicating weak field-induced suppression within this model.} 
\label{fig:4}
\end{center}
\end{figure}
\begin{figure*}[t]
\includegraphics[width=0.99\textwidth]{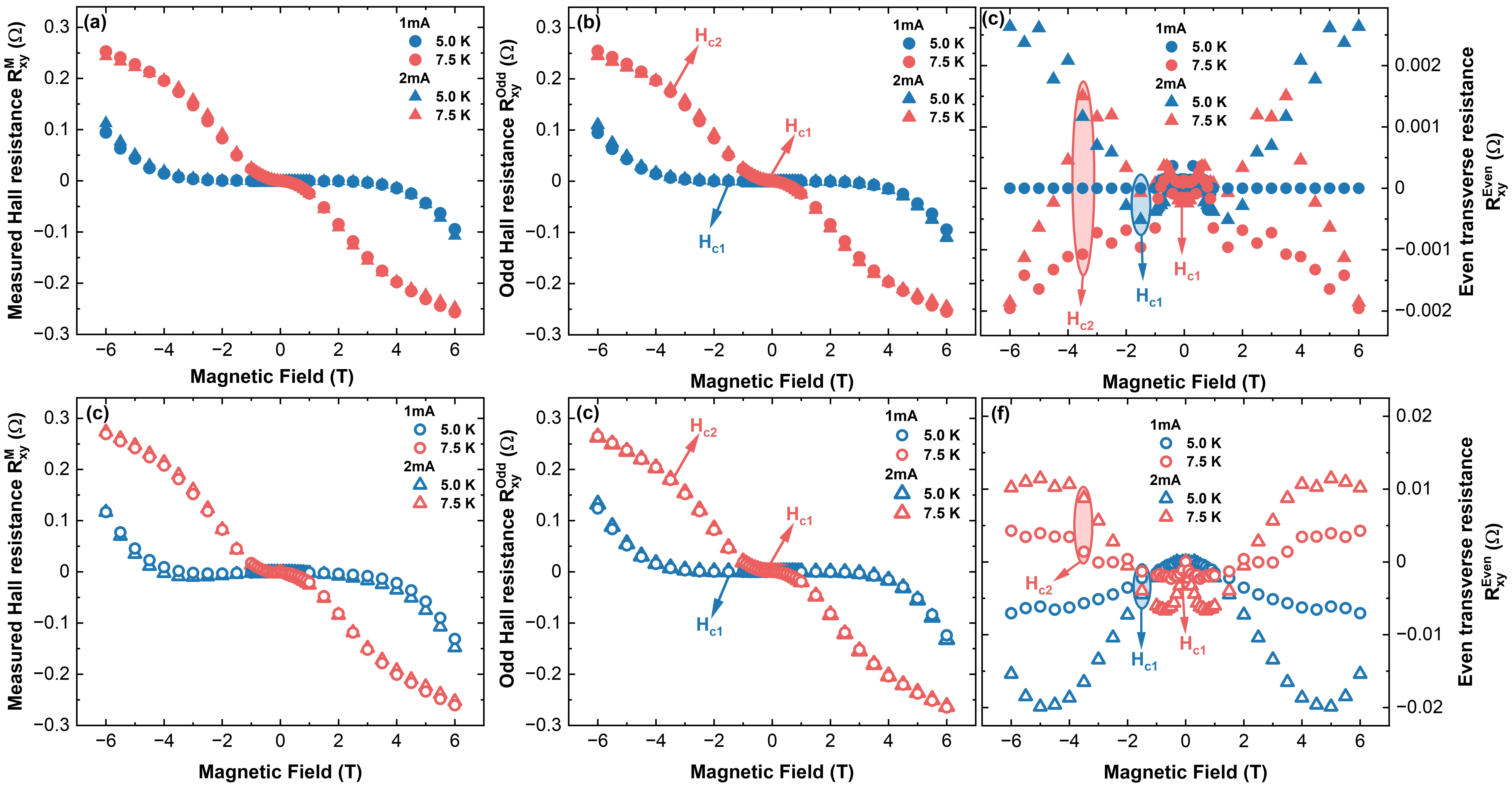}
\caption{Hall resistance, $R^M_{xy}$, and its odd and even field components for 4\,nm thick \textit{NbN} films on Si (a,b,c) and sapphire (d,e,f), measured below $T_{c}$ in Van der Pauw geometry at 5.0\,K and 7.5\,K for applied currents of 1\,mA and 2\,mA. Panels (a,d) correspond to the measured Hall resistance, which shows a nonlinear magnetic-field dependence. The negative slope indicates n\,-type carrier conduction. Panels (b,e) show the odd Hall resistance, $R^{\mathrm{odd}}_{xy}$, obtained by antisymmetrizing $R^M_{xy}$ using Eq.~(\ref{Eq.5}), and panels (c,f) show the even transverse resistance, $R^{\mathrm{even}}_{xy}$, obtained by symmetrizing $R^M_{xy}$ using Eq.~(\ref{Eq.6}). The marked points $H_{c_{1}}$ and $H_{c_{2}}$ indicate the transport characteristic fields corresponding to the onset of measurable dissipation and the crossover toward the normal-state regime, respectively. The even transverse component is enhanced at higher driving current, suggesting a current-dependent transverse response below $T_{c}$.}
\label{fig:5}
\end{figure*}
Fig.~\ref{fig:4} shows the magnetic field dependence of the superconducting current for \textit{NbN} film on Si and sapphire substrates under both out-of-plane and in-plane configurations. For a fixed temperature, $I_{\Superconducting}$ decreases with the applied magnetic field in the out-of-plane configuration, consistent with enhanced vortex-related dissipation and progressive suppression of superconducting transport. A further reduction is observed with increase in temperature, indicating that the superconducting channel weakens as the system moves closer to the transition region. In contrast, for the in-plane magnetic-field configuration, $I_{\Superconducting}$ remains close to the applied driving current over the entire measured field range for both substrates. This indicates that the superconductivity remains largely preserved under a magnetic field applied parallel to the film plane. This strong contrast between the out-of-plane and in-plane responses indicates that the suppression of superconductivity below T$_{c}$ is strongly dependent on the direction of the applied magnetic field. 
We next examine the Hall resistance below T$_{c}$ to analyse the corresponding transverse transport response in the superconducting state.
\subsection{Transverse magnetotransport $R^M_{xy}$}
To investigate the transverse magnetotransport $R^M_{xy}$ below T$_{c}$, we measured the Hall resistance $R^M_{xy}$ in the Van der Pauw geometry with a magnetic field applied perpendicular to the film plane. In an ordinary Hall effect, the transverse voltage arises from the Lorentz force acting on the charge carriers. Therefore, the Hall response is expected to reverse sign when the magnetic field is reversed, showing a field-antisymmetric nature, or odd symmetry in magnetic field:
\begin{eqnarray}
R^M_{xy}(+H) = -R^M_{xy}(-H)
    \label{Eq.4}
\end{eqnarray}
However, below $T_{c}$, the measured transverse voltage can also contain additional field-symmetric contributions arising from non-Hall transverse effects \cite{article40,article41,article42} or from a small admixture of the longitudinal resistance into the transverse voltage due to geometrical asymmetry or nonuniform current paths \cite{article41}. This artifact is commonly written as $\delta R_{xx}$, where $\delta$ is the geometrical mixing factor. We checked the possible longitudinal-resistance mixing contribution, $\delta R_{xx}$, and a possible field-independent instrumental offset as described in the Appendix. For the present Van der Pauw measurements, these terms were found to be much smaller than $R^M_{xy}$ and did not alter the observed field or current dependence of the transverse response. Therefore, no correction was applied to the Van der Pauw data. For the Hall-bar comparison, the longitudinal-mixing factor was calibrated above $T_c$ and no separate constant offset was subtracted, as described with Supplementary Eqs.~(S1) and (S2) and Supplementary Fig.~S4. Following this check, the measured Hall resistance is decomposed into field-antisymmetric ($R^{\mathrm{odd}}_{xy}$) and field-symmetric ($R^{\mathrm{even}}_{xy}$) components as follows \cite{article43}:
\begin{eqnarray}
   R_{xy}^{\mathrm{odd}}(H) = 
   \frac{R_{xy}^{\mathrm{M}}(+H)-R_{xy}^{\mathrm{M}}(-H)}{2}
   \label{Eq.5}
\end{eqnarray}

\begin{eqnarray}
   R_{xy}^{\mathrm{even}}(H) = 
   \frac{R_{xy}^{\mathrm{M}}(+H)+R_{xy}^{\mathrm{M}}(-H)}{2}
   \label{Eq.6}
\end{eqnarray}

Fig.~\ref{fig:5}(b) and Fig.~\ref{fig:5}(e) show the odd component that isolates the field\,-antisymmetric part of the transverse response and therefore represents the Hall-like contribution below T$_{c}$. For NbN on both substrates, the odd Hall signal shows a negative slope, consistent with electron-like carrier conduction, but its field dependence is nonlinear. This shows that transverse transport in the superconducting regime is influenced by the coexistence of superconducting and dissipative contributions rather than by a purely normal-state Hall response. The previously defined $H_{c_{1}}$ and $H_{c_{2}}$ values from the longitudinal magnetotransport are used here as reference fields to discuss the evolution of $R^{\mathrm{odd}}_{xy}$. The strongest change in the odd Hall response occurs within this mixed-state transport window, where superconducting and dissipative contributions coexist. Therefore, $R^{\mathrm{odd}}_{xy}$ is used as a starting point to isolate the normal-conducting Hall resistance by accounting for the superconducting-current contribution, from which the corresponding fluxoid\,-related transport parameters are derived. 
\begin{figure*}[t]
\includegraphics[width=0.99\textwidth]{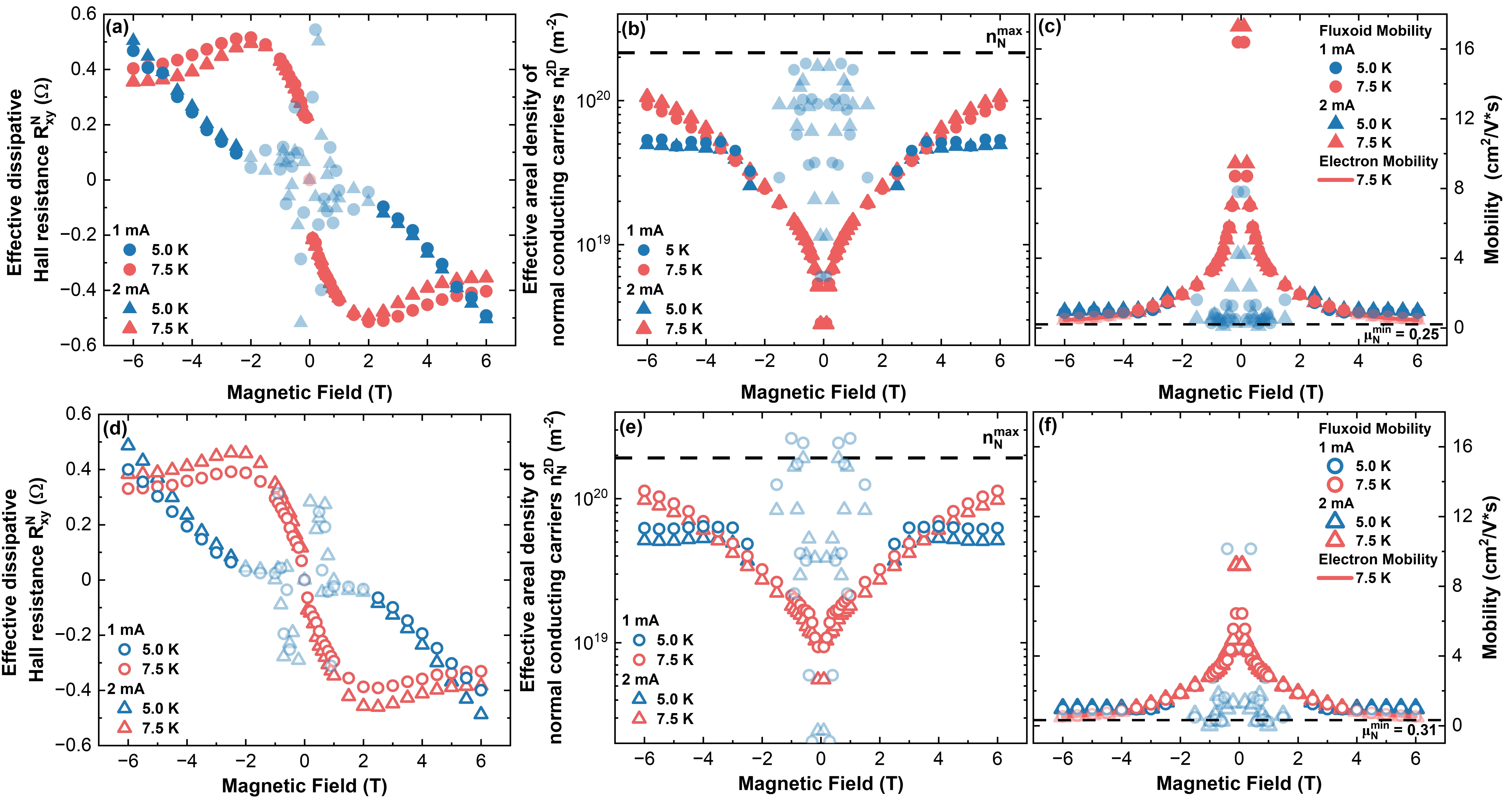}
\caption{(a,d) Derived effective dissipative Hall resistance, $R^{N}_{xy}$, using Eq.~(\ref{Eq.9}) in Van der Pauw geometry at 5.0\,K and 7.5\,K for 4\,nm thick \textit{NbN} films on Si (closed symbols) and sapphire (open symbols). (b,e) Derived effective areal density of dissipative carriers, $n^{2D}_{N}$, using Eqs.~(\ref{Eq.11}) and (\ref{Eq.12}) as a function of magnetic field. (c,f) Model-derived effective dissipative mobility, $\mu$, using Eq.~(\ref{Eq.13}) as a function of magnetic field. The applied currents were 1\,mA and 2\,mA. Faded markers show points that are not used for quantitative interpretation. The label ``Fluxoid Mobility'' in the panel artwork denotes this model-derived quantity and not an independent measurement of vortex velocity.}
\label{fig:6}
\end{figure*}
Since $R^{\mathrm{odd}}_{xy}$ [see Fig.~\ref{fig:5}(b) and Fig.~\ref{fig:5}(e)] preserves the Hall-like contribution below $T_{c}$, it is used for the quantitative analysis of the dissipative normal-conducting channel. In the mixed-state regime between $H_{c_{1}}$ and $H_{c_{2}}$, fluxoids enter the superconducting film and their motion gives rise to dissipative transport. In this region, only a fraction of the driving current contributes to dissipative Hall transport because a finite fraction still flows through the superconducting part. Therefore, to isolate the Hall response associated with the dissipative part, we normalize $R^{\mathrm{odd}}_{xy}$ by the dissipative-current fraction. The effective dissipative Hall resistance is defined as: 
\begin{eqnarray}
R^{N}_{xy}(H) = \frac{R^{\mathrm{odd}}_{xy}(H,T)}{\left(1-\frac{I_{\Superconducting}(H,T)}{I_{\Driving}}\right)}\,.
    \label{Eq.9}
\end{eqnarray}
Here, the denominator represents the fraction of the driving current that contributes to dissipative transport. 

Fig.~\ref{fig:6}(a) and Fig.~\ref{fig:6}(d) shows the magnetic field dependence of the effective dissipative Hall reistance for NbN on two different substrates. For the present below $T_{c}$ measurements, we restrict the quantitative analysis to the mixed state and higher field regions. Below $H_{c_{1}}$, the superconducting current remains close to the driving current, so the denominator in {Eq.(\ref{Eq.9})} becomes very small. In this limit, the extracted $R^{N}_{xy}$ becomes highly sensitive to small voltage offsets and noise. Therefore, the low field points below $H_{c_{1}}$ are not used for quantitative interpretation. Using this effective dissipative Hall resistance, we estimate the corresponding effective dissipative areal carrier density which represents the areal density of carriers involving in dissipative Hall transport in the mixed superconducting state as follows:
\begin{eqnarray}
  n^{2D}_{N}(H,T) = \frac{-H}{R^{N}_{xy}(H,T)\times e},
    \label{Eq.11}
\end{eqnarray}
where e is the charge of an electron. 
At lower temperature  5\,K, (see Fig.~\ref{fig:6}(a,d)) $R^{N}_{xy}$ shows a small magnetic field axis offset. Although fluxoid entry begins near $H_{c_{1}}$, the corrected $R^{N}_{xy}$ doesn't immediately show a stable Hall like response. A linear dissipative Hall behavior is observed only above an onset field ($H_{on}$) around $\pm2.0\,T$. Since the central region between $\approx$  (-2.0\,T to 2.0\,T) contains scattered points and corresponds to weak dissipative transport, it is not used for the areal density analysis. For 5\,K temperature, we define an effective field measured from the onset of the stable dissipative Hall part as $H_{eff}$ $=|H| - H_{on}$, from this the effective areal density of dissipative carriers is estimated as : 
\begin{eqnarray}
  n^{2D}_{\mathrm{N}}(H,T) =
  \left|
  \frac{H_{\mathrm{eff}}}{R^{\mathrm{N}}_{xy}(H,T)\times e}
  \right|,
  \label{Eq.12}
\end{eqnarray}
Fig.~\ref{fig:6}(b) and Fig.~\ref{fig:6}(e) shows the two-dimensional effective dissipative carrier density. At a particular temperature, $n^{2D}_{N}$ shows no pronounced dependence to the change in driving current, indicating that the odd-Hall-derived carrier density is mainly controlled by magnetic field and temperature. At 7.5\,K, $n^{2D}_{N}$ increases with field and approaches the normal-state reference ($n^{max}_{\Normalconducting}$), which is approximately 2.15\,$\times$\,10$^{20}$\,m$^{-2}$ for NbN on Si and 1.90\,$\times$\,10$^{20}$\,m$^{-2}$ for NbN on sapphire. In contrast, at 5\,K, after defining the effective field from the onset of the stable high-field dissipative resistance, $n^{2D}_{N}$ initially increases with field but then remains nearly constant and stays below $n^{max}_{\Normalconducting}$, indicating that the dissipative Hall channel does not fully reach the normal-state reference within the measured field range. This behavior can be compared with the earlier analysis around $T_{c}$ \cite{article28}, where the normal conducting carrier density increased with both magnetic field and temperature due to the growth of the dissipative contribution as superconductivity was weakened. The results below $T_{c}$ follow the same trend for 7.5\,K, but at 5.0\,K the increase is limited, reflecting a stronger superconducting state and incomplete recovery of the normal-conducting channel within the available field range.

We further evaluate the corresponding effective mobility which provides a measure of how efficiently the dissipative carriers associated with the mixed-state Hall response contribute to transport. The mobility is calculated using the same transport relation used in the around $T_{c}$ analysis \cite{article28}. Using the relation $\sigma=1/\rho=ne\mu$, where $n$ is the carrier concentration, $e$ is the electron charge, and $\mu$ is the mobility, we determine $\sigma^{ref}_{N}$ from Eq.~(\ref{Eq.2}). The calculation was carried out using the following relation:
\begin{eqnarray}
\mu = \frac{\sigma^{ref}_{N}\times t}{n_v\times (e_v\times e)}\,
\label{Eq.13}
\end{eqnarray}
where $n_v$ is the areal vortex density, $t$ is the film thickness, and $e_v$ is the model-derived number of dissipative carriers per vortex. In this parametrization, $n_v e_v=n^{2D}_{N}$; therefore Eq.~(\ref{Eq.13}) gives a Hall-derived effective carrier mobility rather than a direct measurement of vortex mobility. The model-derived mobility varies with magnetic field below $T_{c}$ [see Fig.~\ref{fig:6}(c,f)] and is around 0.3\,cm$^2$/Vs at higher fields. The normal-state carrier mobility reported above $T_{c}$ was around 0.26--0.28\,cm$^2$/Vs \cite{article35}. The similar scale does not by itself identify the microscopic carrier or vortex dynamics below $T_c$.

Fig.~\ref{fig:5}(c) and Fig.~\ref{fig:5}(f) show the field\,-symmetric part of the transverse response $R^{\mathrm{even}}_{xy}$ below T$_{c}$. Unlike $R^{\mathrm{odd}}_{xy}$, this component does not change sign with magnetic-field reversal and therefore cannot be assigned to the conventional Hall response. As discussed above, possible longitudinal-resistance mixing was checked through the $\delta R_{xx}$ contribution and is treated in the Appendix. The Hall-bar correction is documented in Supplementary Eqs.~(S1) and (S2) and Supplementary Fig.~S4.

For NbN on both substrates, $R^{\mathrm{even}}_{xy}$ becomes more pronounced at higher driving current and remains finite over the measured field range. In contrast, $\rho^M_{xx}$, $R^M_{xy}$, and $R^{\mathrm{odd}}_{xy}$ remain nearly unchanged with applied current. This indicates that the current-dependent response is mainly associated with the even transverse component rather than with the longitudinal or Hall-like transport channels. In addition to its current dependence, $R^{\mathrm{even}}_{xy}$ also shows a noticeable change between Si and sapphire substrates. The corrected even response also differs among the Hall-bar devices shown in Supplementary Fig.~S5. Because geometry, thickness, temperature, and substrate are not varied independently, the present measurements do not isolate a separate geometry dependence or identify a unique microscopic origin.

We next examine the field evolution of $R^{\mathrm{even}}_{xy}$ using the previously defined transport characteristic fields identified from longitudinal magnetotransport. At H\,=\,0, $R^{\mathrm{even}}_{xy}$ is nearly zero, indicating the absence of measurable symmetric transverse voltage at zero applied field. Upon increasing the magnetic field, a finite symmetric contribution is already visible below the transport-defined onset field $H_{c_{1}}$, where the longitudinal resistance doesn't yet show a clear field-induced dissipative onset (see Appendix Fig.~\ref{fig:7}). This indicates that the low-field even transverse response is not limited to the mixed-state region and cannot be attributed solely to fluxoid dynamics. Above $H_{c_{1}}$, where the longitudinal response begins to show measurable field-induced dissipation, $R^{\mathrm{even}}_{xy}$ becomes more sensitive to the applied driving current. This suggests that nonlinear superconducting dynamics, including possible vortex-related processes, may contribute to this field window. At higher fields, $R^{\mathrm{even}}_{xy}$ evolves gradually and does not vanish abruptly near the transport-defined $H_{c_2}$. This gradual evolution is consistent with a broad resistive crossover, but the present data do not distinguish that explanation from residual superconducting fluctuations or device-dependent mixing. Similar even-in-field transverse voltages have been discussed in the context of superconducting inhomogeneity and current redistribution \cite{article40,article41,article42}, as well as spatially nonuniform flux-flow channels \cite{article45}. Since the microscopic origin of $R^{\mathrm{even}}_{xy}$ cannot be uniquely identified from the present data, we treat it as an additionally observed experimental field-symmetric transverse response and discuss it only in terms of the observed field, current, and sample dependence.  

The observation that $R^{\mathrm{even}}_{xy}$ retains the same sign for positive and negative magnetic field identifies an even-in-field transverse channel rather than the conventional Hall response. Between $H_{c_1}$ and $H_{c_2}$, Abrikosov vortices establish a spatial magnetic-flux and vector-potential texture within the superconducting condensate \cite{Abrikosov1957,book46}. The Aharonov--Bohm effect demonstrates phase sensitivity to the electromagnetic vector potential \cite{article49}. In a superconductor, the vector potential influences the condensate phase and supercurrent through the gauge-invariant phase relation \cite{book46}; however, it is not itself the measured electrical potential. Increasing the driving current changes the Lorentz force acting on the vortices and can modify their velocity and flux-flow trajectories together with the surrounding Cooper-pair current distribution \cite{article47,KoganNakagawa2021}. We therefore consider the possibility that the vortex-induced magnetic and supercurrent texture contributes to $R^{\mathrm{even}}_{xy}$ when mixed-state dynamics is converted into a transverse voltage by broken spatial symmetry, including anisotropic pinning, superconducting inhomogeneity, nonuniform current flow, or device asymmetry \cite{article40,article45}. This response is not assigned directly to the crystal field. The present measurements cannot determine whether $R^{\mathrm{even}}_{xy}$ originates predominantly from vortex-core quasiparticles, Cooper-pair current redistribution, or their combined electrodynamics \cite{article48}.

\section{Conclusion}

In this work, we investigated the below\,-\,T$_{c}$ magnetotransport of ultrathin NbN films to examine how superconducting and dissipative transport evolve under applied magnetic field. The longitudinal magnetotransport shows a strong magnetic field-orientation dependence. The out-of-plane magnetic field suppresses superconductivity much more effectively than the in-plane field, leading to a larger resistive response. From these measurements, we identified the characteristic fields $H_{c_{1}}$ and $H_{c_{2}}$, which define the mixed-state transport window where superconducting and dissipative contributions coexist. The sheet-resistance-based current-partition approach, previously applied around T$_{c}$ \cite{article28}, provides a useful framework for estimating the superconducting-current contribution below T$_{c}$. This analysis shows that the superconducting-current contribution decreases progressively with increasing magnetic field.

We also analyzed the transverse magnetotransport by decomposing the measured Hall resistance into odd and even magnetic-field components. The odd component, $R^{\mathrm{odd}}_{xy}$, retains the Hall-like response and was used to estimate the effective dissipative Hall resistance, Hall-derived dissipative carrier density, and mobility below T$_{c}$. Possible longitudinal-resistance mixing was checked and found to be negligible for the Van der Pauw measurements, while the Hall-bar correction is given in the Supplementary Material. Overall, the results extend the NbN magnetotransport framework from the around-T$_{c}$ regime to the superconducting region below T$_{c}$ and provide a practical way to compare superconducting, dissipative, and Hall-related transport contributions in NbN films.

The even component, $R^{\mathrm{even}}_{xy}$, has the same sign for positive and negative magnetic field, while its magnitude changes with the driving current and differs among the measured samples and devices. A possible interpretation is that Abrikosov vortices create a magnetic-flux and supercurrent texture in the mixed state, and that increasing current modifies vortex motion together with Cooper-pair flow. Coupling of this mixed-state dynamics to an anisotropic pinning landscape, material nonuniformity, or device asymmetry may then produce a measurable even transverse voltage \cite{Abrikosov1957,article40,article45,article47,KoganNakagawa2021}. The current dependence supports a dynamical mixed-state contribution but does not prove that the vortex lattice alone is responsible. The present data do not resolve whether the response originates from vortices, Cooper-pair current redistribution, or a combined contribution from both channels. Distinguishing these possibilities will require current-reversal, angular, reciprocity, temperature-dependent, and vortex-sensitive measurements.

\begin{acknowledgments}
We thank Sven Linzen for helping to prepare the \textit{NbN} samples. Sahitya V. Vegesna acknowledges funding from the Federal Ministry of Education and Research, APPA Photion (Code: 074023-02 BMBF), and from the Office of Technology Assessment at the German Bundestag, TAB-AIF QHub (Code: K203178), and from the Carl-Zeiss-Stiftung (CZS Center QPhoton). Sai Vaishnavi Kanduri acknowledges funding from DFG-IRTG META-Active (Code: 437527638).
\end{acknowledgments}

\appendix*

\section{}
\begin{figure}
\begin{center}
\includegraphics[width=8.6cm]{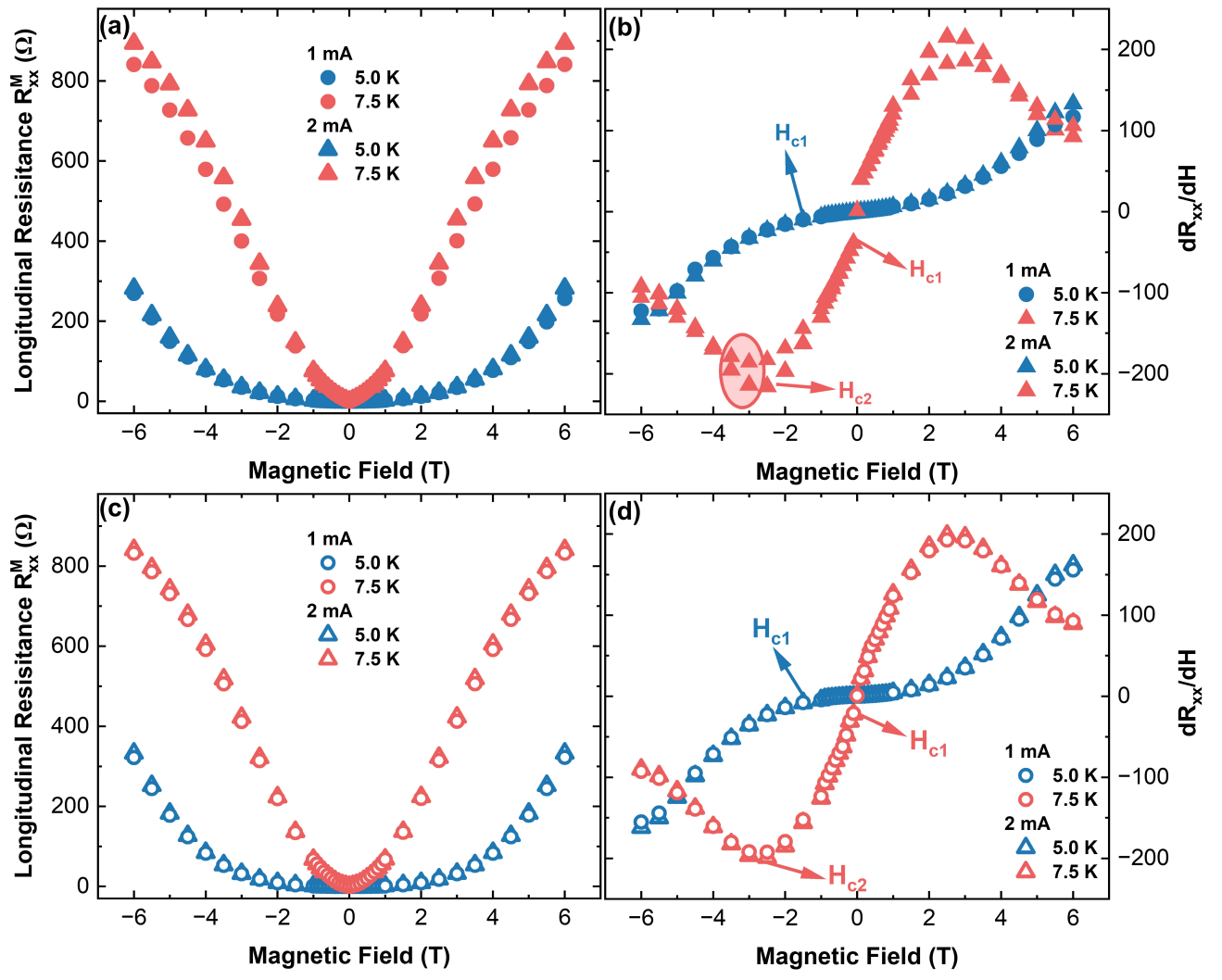}
\caption{(a,c) Longitudinal sheet resistance, $R^M_{xx}$, for a 4\,nm thick \textit{NbN} film on Si (closed symbols) and sapphire (open symbols) substrates, measured below $T_{c}$ in Van der Pauw geometry at temperatures 5.0\,K and 7.5\,K for applied currents of 1\,mA and 2\,mA. Panels (b,d) shows corresponding field derivative of $R^M_{xx}$. The marked points $H_{c_{1}}$ and $H_{c_{2}}$ indicate the transport characteristic fields identified from slope changes, corresponding to the onset of measurable dissipation and the crossover toward the normal-state regime, respectively.}
\label{fig:7}
\end{center}
\end{figure} 

Fig.~\ref{fig:7}(a) and Fig.~\ref{fig:7}(c) show the magnetic-field dependence of the measured longitudinal sheet resistance, $R^M_{xx}$, for NbN films on different substrates. The corresponding field derivatives are shown in Fig.~\ref{fig:7}(b) and Fig.~\ref{fig:7}(d). The extracted transport characteristic fields are summarized in Table~\ref{tab:Hc_values}.

\begin{table*}[t]
\caption{Transport-defined characteristic fields extracted from the derivative analysis. $H_{c_1}$ denotes the onset of a measurable slope change and $H_{c_2}$ the 50\% normal-resistivity crossover. The same reported values were used for both drive currents. No formal uncertainty is available; half the local field interval is approximately $\pm0.05$\,T on the 0.1\,T grid and $\pm0.25$\,T on the 0.5\,T grid, as a sampling-resolution guide only.}
\label{tab:Hc_values}
\begin{ruledtabular}
\begin{tabular}{lcccc}
Substrate & $T$ (K) & $I_D$ (mA) & $|H_{c_1}|$ (T) & $|H_{c_2}|$ (T) \\
\hline
Si & 5.0 & 1 & 1.49 & $\geq6$ \\
Si & 5.0 & 2 & 1.49 & $\geq6$ \\
Si & 7.5 & 1 & 0.09 & 3.49 \\
Si & 7.5 & 2 & 0.09 & 3.49 \\
Sapphire & 5.0 & 1 & 1.49 & $\geq6$ \\
Sapphire & 5.0 & 2 & 1.49 & $\geq6$ \\
Sapphire & 7.5 & 1 & 0.09 & 2--3 \\
Sapphire & 7.5 & 2 & 0.09 & 2--3 \\
\end{tabular}
\end{ruledtabular}
\end{table*}

The sampling-resolution values in Table~\ref{tab:Hc_values} are not complete experimental uncertainties and do not include the choice of criterion, interpolation, field calibration, hysteresis, or sample-dependent systematic effects. At 5\,K, the upper crossover is not reached within the measured $|H|\leq6$\,T range and is therefore reported only as the one-sided bound $|H_{c_2}|\geq6$\,T.

In transverse resistance measurements, a small fraction of the longitudinal voltage can appear in the measured Hall channel because of geometrical asymmetry or nonuniform current paths. This contribution is even in magnetic field and is commonly described as a longitudinal-resistance mixing term, $\delta R_{xx}$, where $\delta$ is a geometrical mixing factor \cite{article41}. Following this approach, the measured transverse resistance can be written as the sum of the intrinsic transverse response, the longitudinal-resistance admixture, and a field-independent offset:
\setcounter{equation}{0}
\renewcommand{\theequation}{A\arabic{equation}}
\begin{eqnarray}
R^{M}_{xy}(H) = R^{\mathrm{true}}_{xy}(H) + \delta R^{M}_{xx}(H) + C ,
\label{Eq.A1}
\end{eqnarray}
Here, C denotes the field-independent voltage offset. The corrected transverse resistance is therefore obtained as
\begin{eqnarray}
R^{\mathrm{corr}}_{xy}(H) = R^{M}_{xy}(H) - \delta R^{M}_{xx}(H) - C ,
\label{Eq.A2}
\end{eqnarray}

The odd and even components can then be obtained from $R^{\mathrm{corr}}_{xy}$ using the same field-symmetrization procedure described in the main text using Eq.~(\ref{Eq.5}) and Eq.~(\ref{Eq.6}). For the Van der Pauw data used in the main analysis, the estimated $\delta R_{xx}$ term was much smaller than the measured transverse response and did not change its field or current dependence; no correction was therefore applied. For the Hall-bar comparison in the Supplementary Material, $R_{xx}$ is explicitly defined as the sheet resistance, $\delta$ is calibrated above $T_c$, and $C=0$ is adopted because the analysis does not document an independent constant offset. The 0.63\,$\Omega$ and 0.39\,$\Omega$ values are the normal-state values of $\delta R_{xx}$ and are therefore not subtracted a second time as $C$.

\bibliography{name.bib}
\clearpage
\onecolumngrid
\raggedbottom

\setcounter{figure}{0}
\setcounter{table}{0}
\setcounter{equation}{0}
\setcounter{section}{0}
\renewcommand{\thefigure}{S\arabic{figure}}
\renewcommand{\thetable}{S\arabic{table}}
\renewcommand{\theequation}{S\arabic{equation}}
\renewcommand{\thesection}{\Roman{section}}
\renewcommand{\theHfigure}{supp.figure.\arabic{figure}}
\renewcommand{\theHtable}{supp.table.\arabic{table}}
\renewcommand{\theHequation}{supp.equation.\arabic{equation}}
\renewcommand{\theHsection}{supp.section.\arabic{section}}

\renewcommand{\section}[1]{%
  \par\addvspace{0.8cm plus 1ex minus 0.2ex}%
  \refstepcounter{section}%
  \begingroup
    \normalfont\small\bfseries\centering
    \MakeUppercase{\Roman{section}.\quad #1}\par
  \endgroup
  \nobreak\vspace{0.5cm}%
}

\begin{center}
{\large\bfseries Mixed\,-State Magnetotransport Below T$_c$ in Type\,-II Superconducting NbN Thin Films\\[0.25em]
-- Supplementary Material --}\par
\vspace{0.9em}
Sai Vaishnavi Kanduri,$^{1,2}$ Danilo Bürger,$^{3}$ Zichao Li,$^{4}$ Mario Ziegler,$^{1}$\\
Detlef Born,$^{1}$ Sahitya V. Vegesna,$^{1,2}$ and Heidemarie Krüger$^{1,2}$\\[0.45em]
{\small\itshape
$^{1}$Leibniz Institute of Photonic Technology, 07745 Jena, Germany\\
$^{2}$Institute for Solid State Physics, Friedrich Schiller University Jena, 07743 Jena, Germany\\
$^{3}$Helmholtz-Zentrum Dresden-Rossendorf (HZDR) Innovation GmbH,\\
Bautzner Landstraße 400, 01328 Dresden, Germany\\
$^{4}$Institute of Ion-Beam Physics and Materials Research,\\
Helmholtz-Zentrum Dresden-Rossendorf (HZDR), Bautzner Landstraße 400, 01328 Dresden, Germany}
\end{center}
\vspace{1.0em}


This Supplementary Material presents the supporting Hall-bar measurements cited in the main manuscript. The figures follow the main-text analysis sequence: longitudinal magnetotransport (Figs.~S1 and S2), the current-partition result (Fig.~S3), the Hall-bar transverse correction and field-parity result (Figs.~S4 and S5), and the effective dissipative parameters (Fig.~S6).

\section{Hall-bar samples and longitudinal magnetotransport}

The Hall-bar sample preparation, naming, transition temperatures, and measurement conditions are given in Table~I and the Experimental Details of the main manuscript. Figures~S1 and S2 present the additional longitudinal-transport data used for the Hall-bar comparison.

\begin{figure}[!htbp]
\begin{center}
\includegraphics[width=0.82\textwidth]{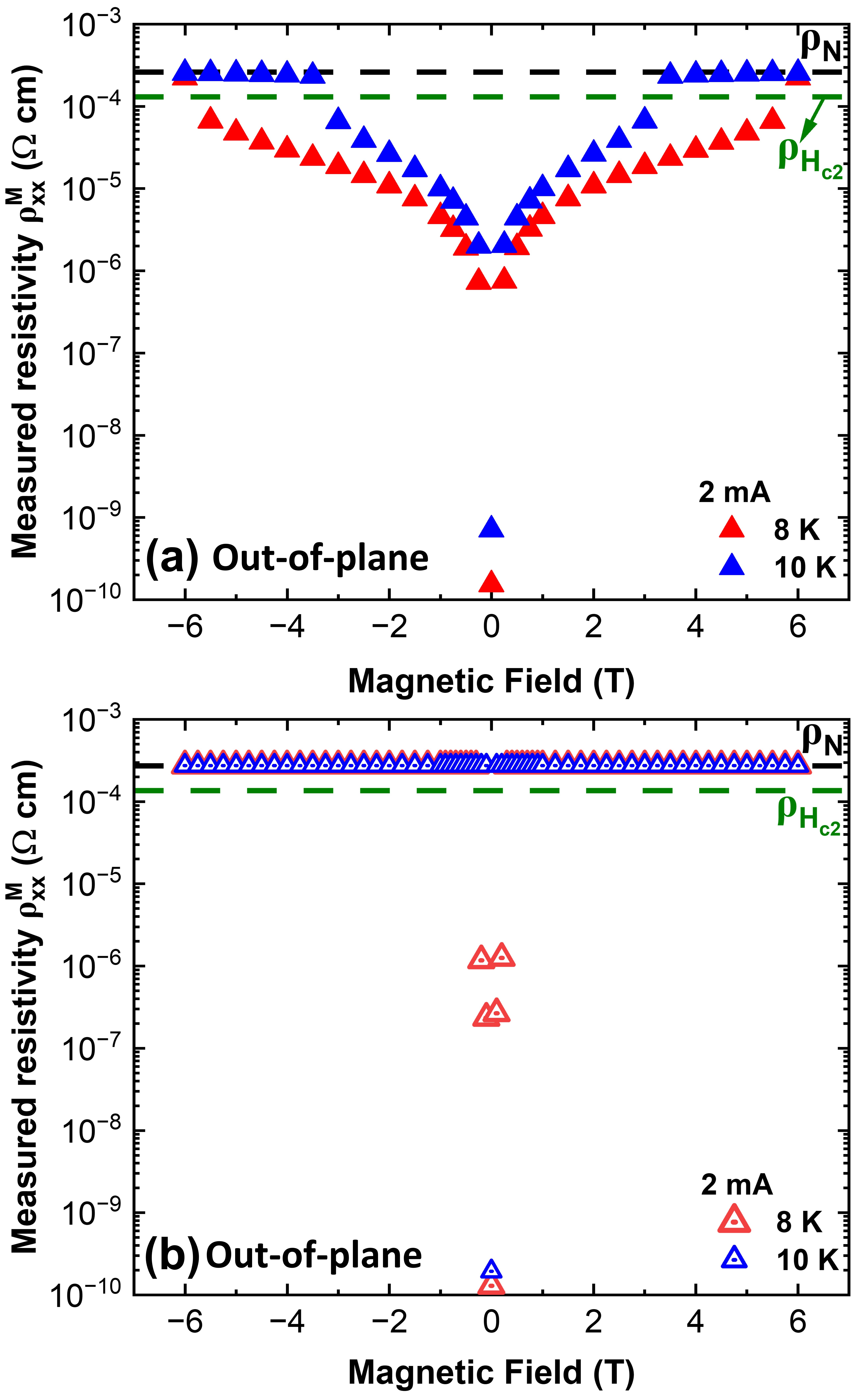}
\caption{Longitudinal resistivity $\rho^M_{xx}$ as a function of out-of-plane magnetic field for (a) the 25\,nm NbN/Si Hall bar and (b) the 25\,nm NbN/glass Hall bar at 8\,K and 10\,K. Measurements were performed from $-6$\,T to $+6$\,T with $I_D=2$\,mA. Black dashed lines show the 50\,K normal-state reference resistivities of approximately 262 and 272\,$\mu\Omega\,\mathrm{cm}$, respectively. Green dashed lines mark one-half of each reference resistivity.}
\label{fig:S1}
\end{center}
\end{figure}

The longitudinal resistance of the 25\,nm NbN/Si Hall bar increases with field magnitude. For the 25\,nm NbN/glass Hall bar, the resistance varies strongly only within a narrow region near zero field and is nearly constant over the remainder of the measured range. The derivative trace for the glass sample therefore does not resolve separate transport characteristic fields.

\begin{figure}[!htbp]
\begin{center}
\includegraphics[width=0.88\textwidth]{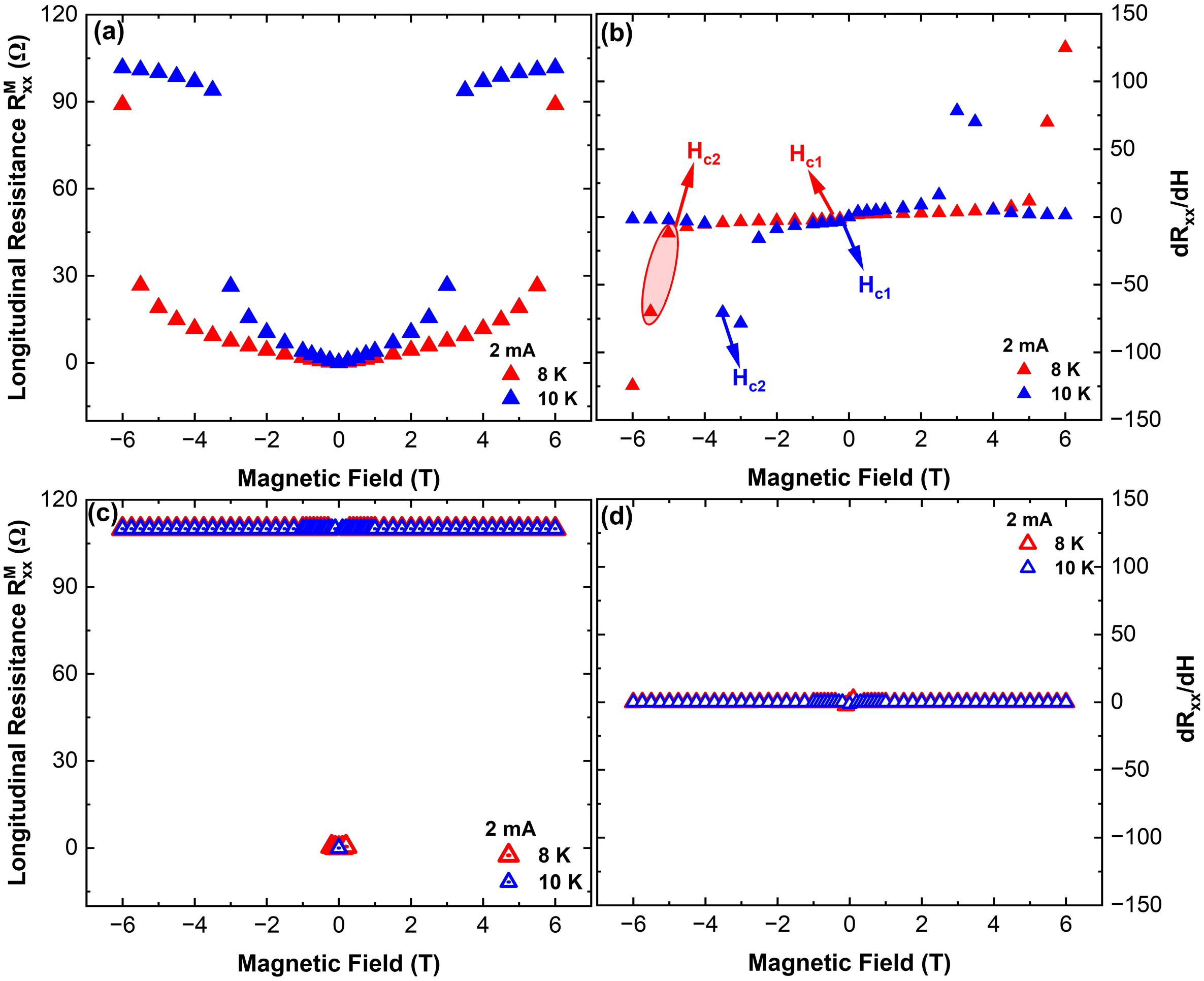}
\caption{(a,c) Longitudinal sheet resistance $R^M_{xx}$ and (b,d) the corresponding field derivatives for (a,b) 25\,nm NbN/Si (closed symbols) and (c,d) 25\,nm NbN/glass (open symbols) at 8\,K and 10\,K with $I_D=2$\,mA. For NbN/glass, the narrow low-field transition does not yield well-separated derivative features that support independent assignments of $H_{c_1}$ and $H_{c_2}$.}
\label{fig:S2}
\end{center}
\end{figure}

Using the 50\,K references shown in Fig.~\ref{fig:S1} and the criterion $\rho^M_{xx}/\rho_N=0.5$, the 25\,nm NbN/Si sample gives crossover fields of 5.77\,T at 8\,K and 3.27\,T at 10\,K. Applying the same criterion to the narrow low-field transition of NbN/glass gives apparent crossings of 0.28\,T at 8\,K and 0.09\,T at 10\,K. These values are criterion-defined crossings and should not be interpreted as independently resolved critical fields.

No formal uncertainty is available for these crossover fields. As a sampling-resolution guide only, half the local field interval is approximately $\pm0.25$\,T for the 25\,nm NbN/Si crossings and $\pm0.05$\,T for the apparent 25\,nm NbN/glass crossings. These values are not complete uncertainties and do not include criterion choice, interpolation, field calibration, hysteresis, or sample-dependent systematic effects. For the narrow low-field transition of NbN/glass, the criterion ambiguity is the dominant limitation.

\FloatBarrier
\section{Superconducting-current estimate}

The current-partition relation in Eq.~(3) of the main manuscript was applied to the Hall-bar data using numerical normal-state references of approximately $2.61\times10^{-4}\,\Omega\,\mathrm{cm}$ for 25\,nm NbN/Si and $2.72\times10^{-4}\,\Omega\,\mathrm{cm}$ for 25\,nm NbN/glass. Figure~S3 shows the resulting Hall-bar current estimates; the interpretation and limitations of this phenomenological partition are those stated in the main text.

\begin{figure}[!htbp]
\begin{center}
\includegraphics[width=0.62\textwidth]{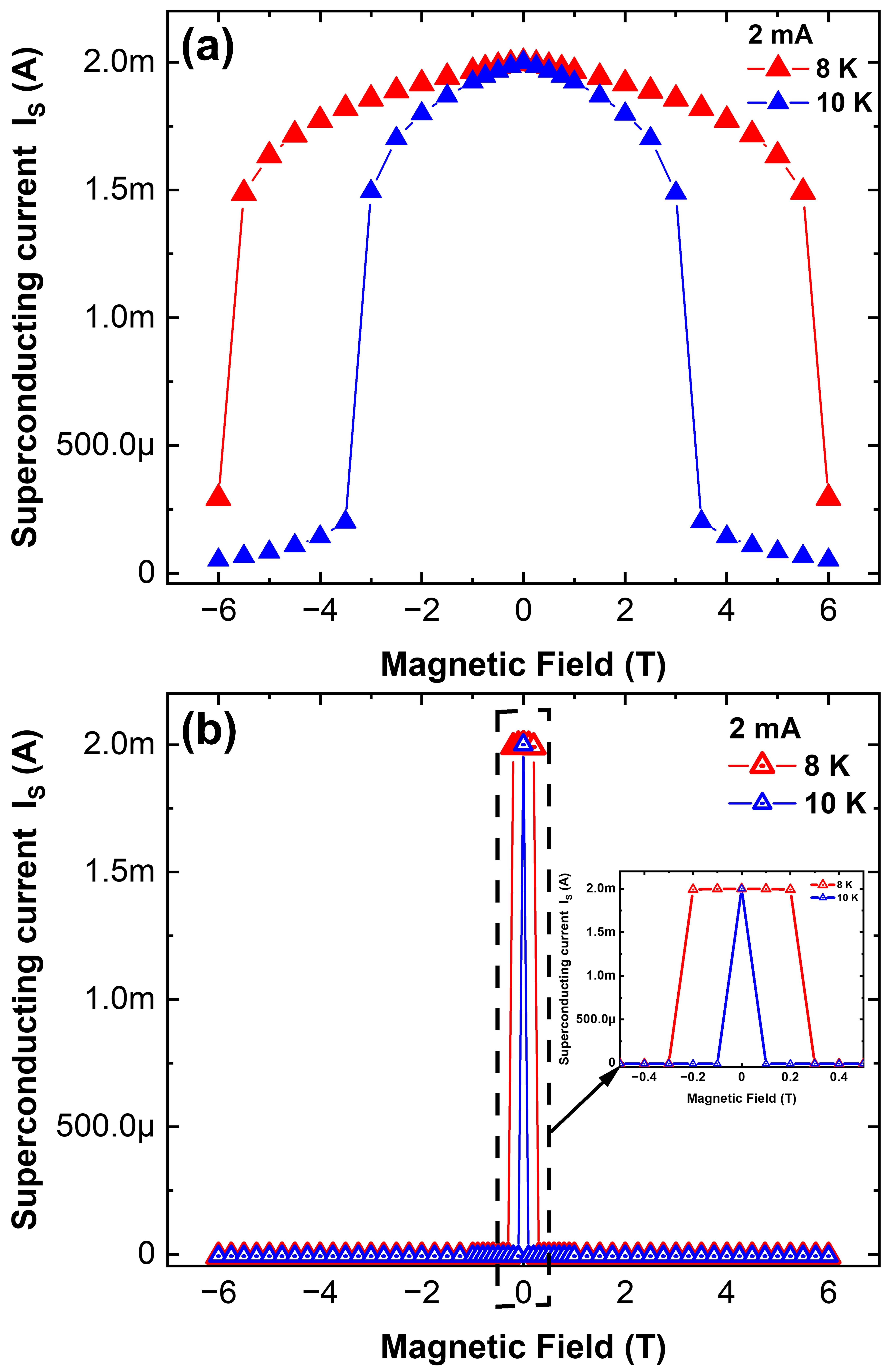}
\caption{Estimated superconducting current $I_S$ from Eq.~(3) of the main manuscript for (a) the 25\,nm NbN/Si Hall bar and (b) the 25\,nm NbN/glass Hall bar at 8\,K and 10\,K with $I_D=2$\,mA. The out-of-plane magnetic field ranges from $-6$\,T to $+6$\,T. The inset in (b) enlarges the region near zero field.}
\label{fig:S3}
\end{center}
\end{figure}

\FloatBarrier
\section{Transverse-response correction and field parity}

\begin{figure}[!htbp]
\begin{center}
\includegraphics[width=0.88\textwidth]{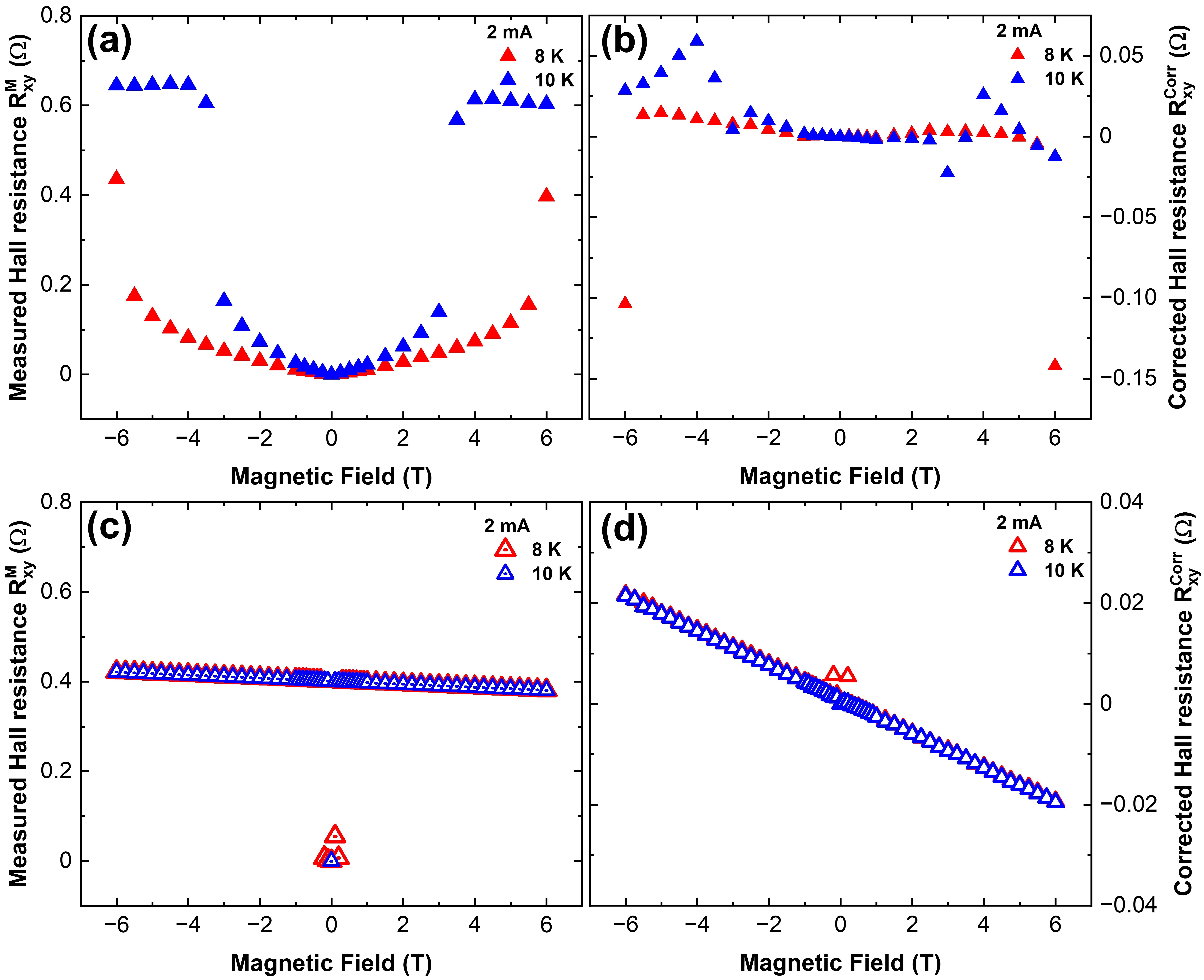}
\caption{Measured and corrected transverse resistance for (a,b) 25\,nm NbN/Si and (c,d) 25\,nm NbN/glass at 8\,K and 10\,K with $I_D=2$\,mA. Panels (a,c) show the measured transverse resistance, and panels (b,d) show the full transverse resistance after subtraction of the longitudinal admixture in Eq.~\ref{Eq:S1}.}
\label{fig:S4}
\end{center}
\end{figure}

The Hall-bar transverse resistance was decomposed using Eqs.~(5) and (6) of the main manuscript. The additional Hall-bar-specific correction in Fig.~\ref{fig:S4} removes longitudinal admixture using the sheet resistance $R^{\mathrm{sheet}}_{xx}=\rho^M_{xx}/t$ according to \cite{article41},
\begin{equation}
R^{\mathrm{corr}}_{xy}(H)=R^M_{xy}(H)-\delta_{\mathrm{sheet}}R^{\mathrm{sheet}}_{xx}(H),
\label{Eq:S1}
\end{equation}
where $\delta_{\mathrm{sheet}}$ is the longitudinal-mixing factor defined for the sheet-resistance convention. For the field-symmetric component, the corresponding correction is
\begin{align}
 R^{\mathrm{sheet,even}}_{xx}(H)&=\frac{R^{\mathrm{sheet}}_{xx}(+H)+R^{\mathrm{sheet}}_{xx}(-H)}{2},\notag\\
R^{\mathrm{even,corr}}_{xy}(H)&=R^{\mathrm{even}}_{xy}(H)-\delta_{\mathrm{sheet}}R^{\mathrm{sheet,even}}_{xx}(H).
\label{Eq:S2}
\end{align}
Within this correction model, the longitudinal admixture is treated through its even-in-field part; the field-antisymmetric component is therefore unchanged. The factor $\delta_{\mathrm{sheet}}$ was calibrated above $T_c$ using $\delta_{\mathrm{sheet}}=\langle R^{\mathrm{even}}_{xy}\rangle/\langle R^{\mathrm{sheet}}_{xx}\rangle$, assuming that the intrinsic even transverse response is negligible in the normal state. Averaging the 15--50\,K normal-state data gives $\delta_{\mathrm{sheet}}=6.061\times10^{-3}$ (0.606\%) for 25\,nm NbN/Si and $3.650\times10^{-3}$ (0.365\%) for 25\,nm NbN/glass.

The numerical value of the mixing factor depends on the normalization chosen for the longitudinal channel; Eq.~\ref{Eq:S1} uses the sheet resistance throughout. The small statistical spread of the above-$T_c$ calibration is not a complete experimental uncertainty because the measurement geometry, field pairing, and assumption of a negligible intrinsic even normal-state response introduce additional systematic effects that are not quantified by the present data.

\begin{figure}[!htbp]
\begin{center}
\includegraphics[width=0.88\textwidth]{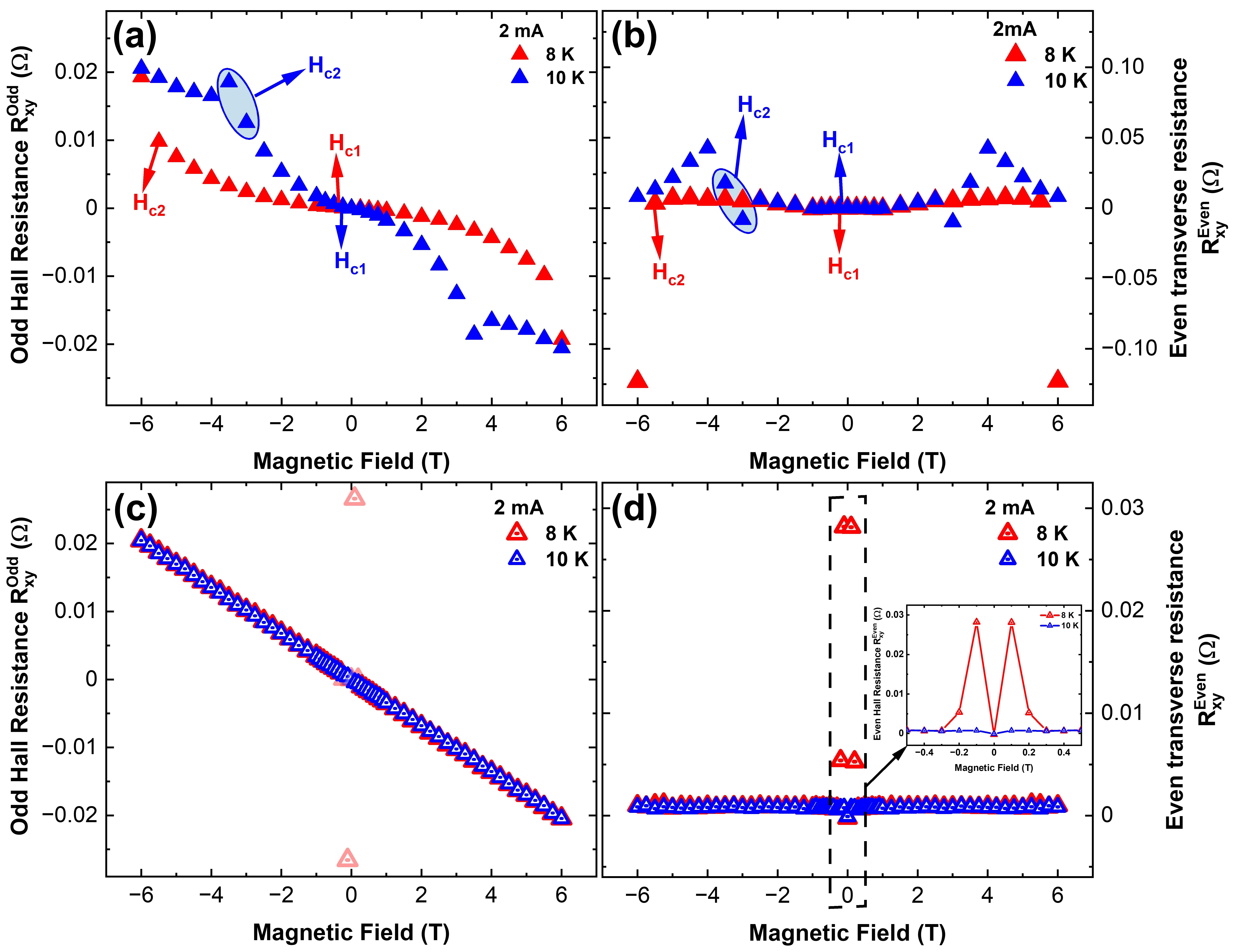}
\caption{Field-antisymmetric transverse resistance $R^{\mathrm{odd}}_{xy}$ in (a,c) and corrected field-symmetric transverse resistance $R^{\mathrm{even,corr}}_{xy}$ in (b,d) for 25\,nm NbN/Si (a,b) and 25\,nm NbN/glass (c,d) at 8\,K and 10\,K with $I_D=2$\,mA. The inset in (d) enlarges the region near zero field. Faded markers identify points excluded from quantitative interpretation; no conclusion in the text relies on these points.}
\label{fig:S5}
\end{center}
\end{figure}

\FloatBarrier
\section{Effective dissipative parameters}

\begin{figure}[!htbp]
\begin{center}
\includegraphics[width=0.98\textwidth]{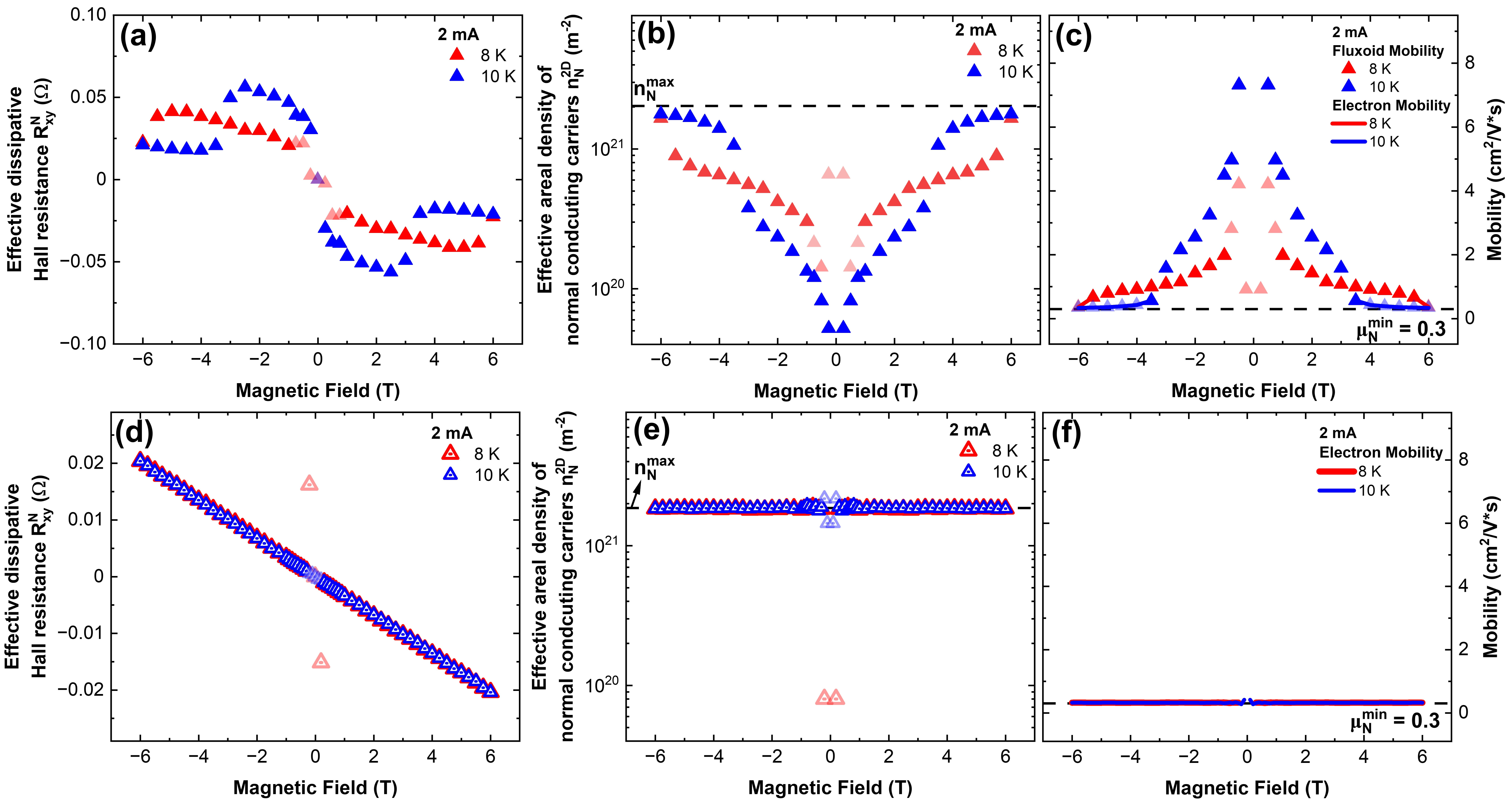}
\caption{Effective dissipative Hall resistance $R^N_{xy}$ in (a,d), Hall-derived effective areal carrier density $n_N^{2D}$ in (b,e), and mobility in (c,f) for 25\,nm NbN/Si (a--c) and 25\,nm NbN/glass (d--f) at 8\,K and 10\,K with $I_D=2$\,mA. Panel (c) shows the model-derived effective-mobility markers together with normal-state electron-mobility reference curves; panel (f) shows the electron-mobility reference curves. Dashed horizontal lines show normal-state reference values. Faded markers denote excluded points as described in the text and are not interpreted quantitatively.}
\label{fig:S6}
\end{center}
\end{figure}

The quantities in Fig.~\ref{fig:S6} were obtained by applying Eqs.~(9) and (11) of the main manuscript to the Hall-bar data. Points affected by the ill-conditioned low-dissipative-current limit described after main-text Eq.~(9), together with other points for which no stable trend was obtained, are shown as faded markers in Figs.~\ref{fig:S5} and \ref{fig:S6} and are omitted from quantitative interpretation.

The mobility-like quantity plotted as triangles in Fig.~\ref{fig:S6}(c) was evaluated using Eq.~(13) and the model parametrization defined in the main manuscript. The curves in panels (c) and (f) show the corresponding normal-state electron-mobility references.

At 50\,K, the normal-state reference values for 25\,nm NbN/Si are $n^{3D}=8.149\times10^{22}$\,cm$^{-3}$, $n^{2D}=2.037\times10^{21}$\,m$^{-2}$, and $\mu=0.29$\,cm$^2$V$^{-1}$s$^{-1}$. The corresponding values for 25\,nm NbN/glass are $n^{3D}=7.414\times10^{22}$\,cm$^{-3}$, $n^{2D}=1.853\times10^{21}$\,m$^{-2}$, and $\mu=0.30$\,cm$^2$V$^{-1}$s$^{-1}$.

\end{document}